\documentclass[amsmath,amssymb,superscriptaddress,nofootinbib]{revtex4}
\usepackage{graphicx}
\usepackage{aas_macros,braket}
\usepackage{natbib}
\input{colordvi.tex}

\begin{document}
%\draft
%
\title{21cm fluctuations from primordial magnetic fields}
\author{Maresuke Shiraishi}
\email{maresuke.shiraishi@pd.infn.it}
\affiliation{%
Dipartimento di Fisica e Astronomia ``G. Galilei'', Universit\`a degli Studi di Padova, via Marzolo 8, I-35131, Padova, Italy
}
\affiliation{%
INFN, Sezione di Padova, via Marzolo 8, I-35131, Padova, Italy
}
\author{Hiroyuki Tashiro}
\email{hiroyuki.tashiro@asu.edu}
\affiliation{%
Physics Department, Arizona State University, Tempe, Arizona 85287, USA
} 
\author{Kiyotomo Ichiki}
\email{ichiki@a.phys.nagoya-u.ac.jp}
\affiliation{%
Kobayashi-Maskawa Institute for the Origin of Particles and
the Universe, Nagoya University, Chikusa-ku, Nagoya, 464-8602, Japan
}
\affiliation{%
Department of Physics and Astrophysics, Nagoya University, Nagoya
464-8602, Japan
}

\date{\today} \preprint{}
\begin{abstract}
Recent observations of magnetic fields in intergalactic void regions and
 in high redshift galaxies may indicate that large scale magnetic fields
 have a primordial origin. If primordial magnetic fields were present soon after
 the recombination epoch, they would have induced density fluctuations on the
 one hand and dissipated their energy into the primordial gas on the
 other, and thereby significantly alter the thermal history of the
 Universe. Here we consider both the effects
 and calculate the brightness temperature fluctuations of the 21cm line
 using simple Monte Carlo simulations. We find that the fluctuations of the 21cm line from the energy dissipation appear only on very small scales and those
 from the density fluctuations always dominate on observationally
 relevant angular scales.
\end{abstract}
\maketitle

\section{Introduction}

Numerous astronomical observations have suggested that magnetic fields
are ubiquitous in the Universe. They exist not only in galaxies, but
also in even larger systems, such as in clusters of galaxies. The origin
of such large scale magnetic fields is still a matter of debate \cite{2002RvMP...74..775W,2012SSRv..166...37W,2013A&ARv..21...62D}. It is now believed that the magnetohydrodynamics (MHD) dynamo is a very powerful mechanism to amplify and maintain the galactic magnetic fields. However, the dynamo mechanism does not explain the origin of magnetic fields itself. It is shown that the seed fields as large as $10^{-20}$ $\sim$
$10^{-30}$ G are necessary to explain the observed magnetic fields 
of $10^{-6}$ G in galaxies and clusters of galaxies \cite{1999PhRvD..60b1301D}.

Primordial magnetic fields have been intensively studied in the
literature as a possible origin of the large scale magnetic fields.  A
variety of mechanisms to generate the primordial magnetic fields have
been proposed, which include inflation with a break of conformal
invariance
\cite{Ratra:1991bn,Turner:1987bw,2001PhLB..501..165D,2004PhRvD..69d3507B,2008JCAP...01..025M,2012PhRvD..86b3512S,2012JCAP...10..034F},
effects at phase transitions in the early Universe
\cite{1983PhRvL..51.1488H,1991PhLB..265..258V,1998PhRvD..58j3505H,1998PhRvD..57..664A,2010PhRvD..81h5035H,2013PhRvD..87h3007K,2014JCAP...02..036L},
and cosmological vector modes in first and/or second order cosmological
perturbations
\cite{2005PhRvL..95l1301T,2006Sci...311..827I,2009CQGra..26m5014M,2007astro.ph..1329I,2010arXiv1012.2958F,2005PhRvD..71d3502M,1970MNRAS.147..279H,Hogan:2000gv,2005MNRAS.363..521G,2004APh....21...59B,2013PhRvD..87j4025S,2012PhRvD..85d3009I}.
The field strength of generated magnetic fields varies depending on the
proposed models, and the fields are often parametrized by the magnetic
field amplitude normalized at $\lambda=1$ Mpc scale $B_\lambda$ and the
index $n_B$ of the power spectrum of magnetic fields.

Recent observations of magnetic fields in galaxies at high redshift \cite{2008Natur.454..302B,2013arXiv1307.2678J,2013arXiv1305.1450N} and
in void regions (under discussion \cite{2010Sci...328...73N,2010ApJ...722L..39A,2013ApJ...771L..42T,2012ApJ...744L...7T,2011MNRAS.414.3566T}) may support the hypothesis that the seed fields are of
primordial origin. If this is the case, the primordial magnetic fields
have influenced many kinds of cosmological processes, such as the big
bang nucleosynthesis, cosmic microwave background (CMB) anisotropies, and the
formation of large scale structure of the Universe (see~\cite{2012PhR...517..141Y} and references therein). Recently, the Planck Collaboration placed limits on primordial magnetic fields as
$B_{\lambda}<3.4$ nG and $n_B<0$ from the 
temperature anisotropies at large and small angular scales
\cite{2013arXiv1303.5076P}. 

In this paper, we consider the effect of primordial magnetic fields on
cosmological 21cm signals during the dark age. In particular, we investigate the
thermal history of the primordial hydrogen gas in the Universe by taking
into account the heat injection due to the ambipolar diffusion of the
magnetic fields \cite{2005MNRAS.356..778S}. The heat injection from the magnetic fields into the
weakly ionized primordial gas will leave a unique signature in the future
21cm observations. If the heating raises the gas temperature and, hence, the spin temperature becomes high above the background CMB temperature, the 21cm signal comes as an emission even at
redshift $z\gtrsim 20$, while the signal during the dark age is expected to be absorption
in the standard thermal history of the Universe.

The global effects of the heating from primordial magnetic fields have already been studied in Refs.~\cite{2005MNRAS.356..778S,2006MNRAS.368..965T,2006MNRAS.372.1060T,2008PhRvD..78h3005S,2009ApJ...692..236S,
2013arXiv1309.7994K}.  
In particular, Refs.~\cite{2006MNRAS.372.1060T,2009ApJ...692..236S} have
investigated these global effect on the 21cm signals.
They found that the evolution of gas temperature can change
significantly, and the temperature rises as high as $T_g \gtrsim
10^{3}$ K in redshift $10\lesssim z \lesssim 1000$ if nanogauss
magnetic fields (comoving) are considered. Here we extend their
investigations by taking the spatial fluctuations of the heat injection into
account. Since it is considered that primordial magnetic fields are
possibly distributed following Gaussian random statistics, we also
expect that the heating rate also has fluctuations in space and nontrivial
correlations. One complicated thing here is that the correlation will be
highly nonlinear: one has to evaluate
eight-point correlation functions of the primordial magnetic fields because the ambipolar heating term is proportional to
the magnetic field values to the fourth power. In this paper we utilize
simple Monte Carlo simulations to
evaluate the correlation, 
and examine the 
power spectrum of the heating rate and the corresponding 21cm signal from
the high redshift Universe. 
This is partly motivated by the fact that proposed future observations
of HI gas in high redshifts are performed by interferometers~such as the Square Kilometer Array (SKA),
and therefore signals with contrast, i.e., fluctuations,
are easier to detect than the global ones.
Throughout this paper, we adopt the standard $\Lambda$CDM model with $h=0.7$, $
\Omega_b h^2 = 0.0223$, and $\Omega_c h^2 = 0.104$, 
where $h$ is the present Hubble constant normalized by $100~{\rm km / s
/ Mpc}$ and $\Omega_b$ and $\Omega_c$ are the present baryon and cold dark matter (CDM) energy density parameters, respectively.

%%%%%%%%%%%%%%%%%%%%%%%%%%%%%%%%%%%%%%%%%%%%%%%%%%%%%%%%%%

\section{Theory of ambipolar diffusion in the dark age}

After the recombination epoch,
magnetic fields create the difference in motion between neutral and
ionized baryons by their Lorentz force. This velocity difference induces finite viscosity
in the baryon fluid. As a result, the magnetic field energy is transferred to
the thermal energy of baryons. 
This process is the so-called ambipolar diffusion~\cite{1992phas.book.....S}
and can happen at the
late-time Universe in the dark age under the presence of 
electrically neutral particles~\cite{2005MNRAS.356..778S}.

The dissipation of the magnetic field energy heats the intergalactic medium~(IGM) gas, and the
resultant thermal ionization becomes effective. As a result,
cosmological 21cm signals are altered. In this section, we briefly
review the theoretical aspects of the evolution of the IGM gas and 21cm
signals with ambipolar diffusion. For simplicity, we neglect the
existence of helium in the IGM gas throughout this paper.

\subsection{Temperature and ionization evolution of IGM gas}
The energy dissipation rate due to ambipolar diffusion is
given by~\cite{1956MNRAS.116..114C}
\begin{eqnarray}
\Gamma = \frac{|(\nabla \times {\bf B}) \times {\bf B}|^2}{16\pi^2 \chi
 \rho_b^2 x_i} ~,
\label{eq:gamma}
\end{eqnarray}
where $\rho_b$ is the baryon energy density and $\chi = 3.5 \times 10^{13} ~
{\rm cm^3 g^{-1} s^{-1}}$ denotes the drag coefficient~\cite{1983ApJ...264..485D}.

The evolutions of the hydrogen gas temperature $T_g$ are given by~\cite{2005MNRAS.356..778S}
\begin{equation}
 \frac{d T_g }{d t} = - 2 H T_g +\frac{x_i }{ 1+x_i} 
\frac{8 \rho_\gamma \sigma_T }{ 3 m_e c} (T_\gamma -T_g) +
 \frac{\Gamma }{ 1.5 k_B n_H}, \label{eq:Te}
\end{equation}
where $k_B$, $\rho_\gamma$, $H$, $x_i$, $n_H$ $m_e$ and $\sigma_T$ denote
the Boltzmann constant, the photon energy density, the Hubble parameter,
the ionization fraction, the hydrogen number density, the electron mass, and the cross section of Thomson
scattering, respectively. 
The dissipation of primordial magnetic fields after the recombination
epoch is also induced by the nonlinear decaying of MHD modes.
This dissipation is effective around $z > 800$~\cite{2005MNRAS.356..778S,2013arXiv1309.7994K}. Since we are interested
in redshifts $z \sim 20$ which are the observable redshifts for future
21cm observations, we neglect the dissipation due to the MHD decaying.

The evolution of the hydrogen ionization fraction is determined by
photoionization by the CMB, radiative recombination and thermal collisional ionization,
\begin{equation}
 \frac{d  x_i }{ dt}
=   \left [ \beta_e (1-x_i)\exp\left(-\frac{h\nu_\alpha }{ k_BT_\gamma}\right )-\alpha_e n_H x_i^2 \right ] C + \gamma_e n_H (1-x_i)x_i,
\label{eq:xi}
\end{equation}
where $\alpha_e$ is the recombination coefficient, $\beta_e$ is the
photo ionization coefficient, $\gamma_e$ is the collisional
ionization coefficient, and $h \nu_\alpha = 13.6$~eV is the ground
state binding energy of hydrogen. For these coefficients and $C$, we refer to 
Refs.~\cite{1999ApJ...523L...1S,2000ApJS..128..407S}. %RECFAST 

Although primordial magnetic fields are not detected yet, it is generally expected
that the fields are tangled. Therefore, the dissipation rate $\Gamma$
could be spatially fluctuated. The resultant temperature and
ionization fraction also are fluctuated following Eqs.~(\ref{eq:Te}) and (\ref{eq:xi}).

\subsection{21cm signals from IGM gas}

In the Rayleigh-Jeans limit,
the intensity of cosmological 21cm signals at a frequency $\nu$ can be written in terms of
the brightness temperature $T_{21}$,
\begin{equation}
 I(\nu) = \frac{2 \nu^2}{c^2} k_B T_{21}(\nu).
\end{equation}

Since the 21cm signal is observed as an emission or absorption signal
against the CMB, it is useful to define the
differential brightness temperature against the CMB temperature, $\delta
T_{21} = T_{21} - T_\gamma$.
The positive $\delta T_{21}$ represents that the 21cm signal is an
emission against the CMB, while the negative one means that the signal
is an absorption.
For a given frequency, the differential brightness temperature is given by
\begin{equation}
 \delta T_{21} = \frac{T_s(z) - T_\gamma (z) }{1+z}(1- e^{-\tau(z)}),
\label{eq:diff_Tb}
\end{equation}
where $z$ is the redshift corresponding to the frequency of
observation,~$1+z = \nu_{21}/\nu$ with $\nu_{21}$ denoting the frequency of the 21cm line, $T_s(z)$ is the spin temperature, and $\tau(z)$ is the optical depth
of the IGM at $z$.

The optical depth of the IGM to the hyperfine transition is expressed
by~\cite{1997ApJ...475..429M} 
\begin{equation}
 \tau (z) = \frac{3 c^3 \hbar A_{10} n_{\rm HI}}{16 k_B \nu_{21}^2 T_s H(z)},\label{eq:opt}
\end{equation}
where $A_{10}$ is the spontaneous emission coefficient for the
transition, $A_{10} = 2.85 \times  10^{-15} ~{\rm s}^{-1}$, and $n_{\rm HI}$
is the number density of HI gas. The spin temperature represents the ratio of the hyperfine level populations. Since the spin temperature of the IGM is determined by the balance among absorption of CMB photons, thermal collisional
excitation, and Lyman-$\alpha$ pumping~\cite{1952AJ.....57R..31W,1958PIRE...46..240F}, we can obtain the spin
temperature from 
\begin{equation}
 T_s =\frac{T_\gamma + y_k T_g +y_\alpha T_\alpha}{1+y_k + y_\alpha},
\end{equation}
where $T_\alpha $ is the color temperature of Ly-$\alpha$ flux, and
$y_k$ and $y_\alpha$ are the
kinetic and the Ly-$\alpha$ coupling terms, respectively~\cite{1997ApJ...475..429M}.
Because we assume that there are no Ly-$\alpha$ sources such as
stars and galaxies in our interesting redshifts for simplicity, we neglect
the Ly-$\alpha$ coupling term.

As shown in the previous section, the gas temperature induced by the
ambipolar diffusion depends on the fourth power of Gaussian magnetic
fields. Hence, to obtain the power spectrum of the brightness temperature
fluctuations, we have to evaluate an eight-point correlation function of
Gaussian magnetic fields. Since computing it analytically is quite
complicated, we here shall estimate by use of random realizations of
magnetic fields obtained through brute-force Monte Carlo simulations.

%%%%%%%%%%%%%%%%%%%%%%%%%%%%%%%%%%%%%%%%%%%%%%%%%%%%%%%%%%
\section{Monte Carlo simulations}

Let us start from a situation that seed magnetic fields are created in the very early Universe, and adiabatically decay as ${\bf B}({\bf x}, t) = {\bf B}_0 ({\bf x}) /
a^{2}$ with $a(t)$ being the scale factor. We assume that magnetic
fields are stochastically homogeneous and isotropic. Conventionally, their power
spectrum at the present time is parametrized by a strength smoothed on
$\lambda = 1$ Mpc, $B_\lambda$, and a simple power-law function with a
spectral index, $n_B$, as \cite{2010PhRvD..81d3517S}
\begin{eqnarray}
\begin{split}
\Braket{B_{0i}({\bf k}) B_{0j}^{*}({\bf p})} 
&= (2\pi)^3 \frac{P_B(k)}{2} 
\left( \delta_{ij} - \hat{k}_i \hat{k}_j \right)
\delta^{(3)} ({\bf k} - {\bf p}) ~, \\
%-----
P_B(k) &= \frac{\left(2 \pi \right)^{n_B + 5} B_\lambda^2}
{\Gamma(\frac{n_B + 3}{2}) k_\lambda^{n_B + 3} } k^{n_B}~~~\mbox{for}~~~k<k_c~, \label{eq:PMF_def}
\end{split}
\end{eqnarray}
where $k_\lambda = 2\pi / \lambda$ and $k_c$ is the cutoff wave number of magnetic fields.\footnote{The small amplitude of $B_\lambda$ does not mean that the induced effects are small. The total energy density associated with magnetic fields might be large enough; e.g., see Ref.~\cite{Kahniashvili:2010wm}.} 
The cutoff scale is determined by the radiation viscosity
at the recombination epoch \cite{Jedamzik:1996wp,Subramanian:1997gi} and given by 
\begin{equation}
k_{\rm c}=\left[143\left(\frac{B_\lambda}{1~{\rm nG}}\right)^{-1}\left(\frac{h}{0.7}\right)^{1/2}\left(\frac{\Omega_b h^2}{0.021}\right)^{1/2}\right]^{2/(n_B+5)}{\rm Mpc}^{-1},
\end{equation}
in the matter dominated universe. 

Practically, instead of magnetic fields, we simulate vector potentials
${\bf A}({\bf k})$ in $k$ space with $512^3$ grids, whose power spectrum
reads
\begin{eqnarray}
\Braket{A_i({\bf k}) A_j^{*}({\bf p})} 
= (2\pi)^3 \frac{P_B(k)}{2k^2} \delta_{ij} \delta^{(3)}({\bf k} - {\bf p})~.
\end{eqnarray}
Then, every off-diagonal component vanishes and, hence, numerical operations
are reduced drastically in comparison with the direct simulation of
magnetic fields. After that, we convert realizations of the vector
potentials into those of magnetic fields following the definition of
vector potentials in $k$ space as
\begin{eqnarray}
{\bf B}({\bf k}) = i{\bf k} \times {\bf A}({\bf k}) ~.
\end{eqnarray}
With ${\bf B}({\bf k})$ at hand, the calculation for the energy
dissipation rate due to ambipolar diffusion is straightforward: we take
a rotation once again in $k$ space, $i{\bf k} \times {\bf B}({\bf k})$, move
to real space, to get $\nabla \times {\bf B}({\bf x})$, take a vector
product with ${\bf B}({\bf x})$ in real space and obtain the energy
dissipation rate $\Gamma({\bf x})$ in Eq.~(\ref{eq:gamma}).  We then
calculate the thermal history at every pixel for a realization of
$\Gamma({\bf x})$ and estimate the power spectrum of temperature
fluctuations.  

\begin{figure}
%\rotatebox{0}{\includegraphics[width=1.0\textwidth]{TestFigs/fig1.eps}}
\rotatebox{0}{\includegraphics[width=1.0\textwidth]{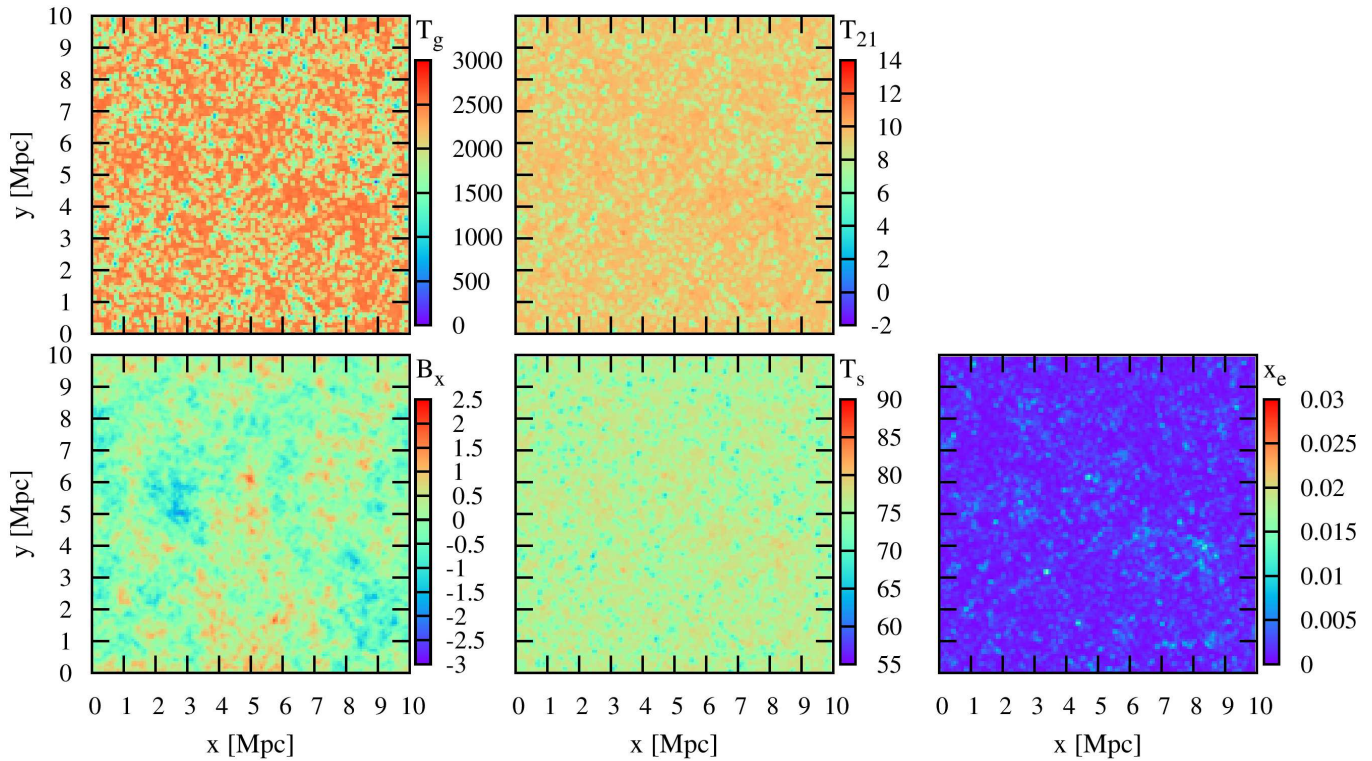}}
\caption{Realization of (the $x$ component of) primordial magnetic
 fields~[nG]~(bottom left) and corresponding fluctuations of the gas temperature~[K]~(top
 left), the spin temperature~[K]~(bottom center), the 21cm brightness 
 temperature~[mK]~(top center), and the ionization fraction (bottom right)
 with parameters $B_\lambda=1$ [nG] and $n_B=-2.9$. The heating is
 dominated by contributions from small scale structures of magnetic fields. It can be seen that
 the spin
 temperature and the ionization fraction fluctuations are positively correlated.}
\label{fig:simulation}
\end{figure}

Figure~\ref{fig:simulation} shows an example of realizations of
(the $x$ component of) primordial magnetic fields $B_x$ together with the
corresponding gas temperature $T_g$, the spin temperature $T_s$,
the 21cm brightness temperature $T_{\rm 21}$ and the ionization fraction
$x_i$. As is evident from the figure, distributions of these temperature
fluctuations are far from the Gaussian distributions even though magnetic
fields are Gaussian distributed. The heating is dominated by the
contribution from small scale
structures of magnetic fields and thus it is important to resolve the
cutoff scale of 
magnetic fields in Monte Carlo realizations.  It should be noted that
the spin temperature and ionization fraction fluctuations are positively
correlated; namely, the region which has larger spin temperature has a 
larger ionization fraction. Consequently, the larger ionization fraction
cancels out in part the contribution from the spin temperature to the
21cm brightness temperature, leading to smoother 21cm brightness
temperature fluctuations.

\begin{figure}
 \begin{minipage}[m]{0.49\textwidth}
\rotatebox{0}{\includegraphics[width=1.0\textwidth]{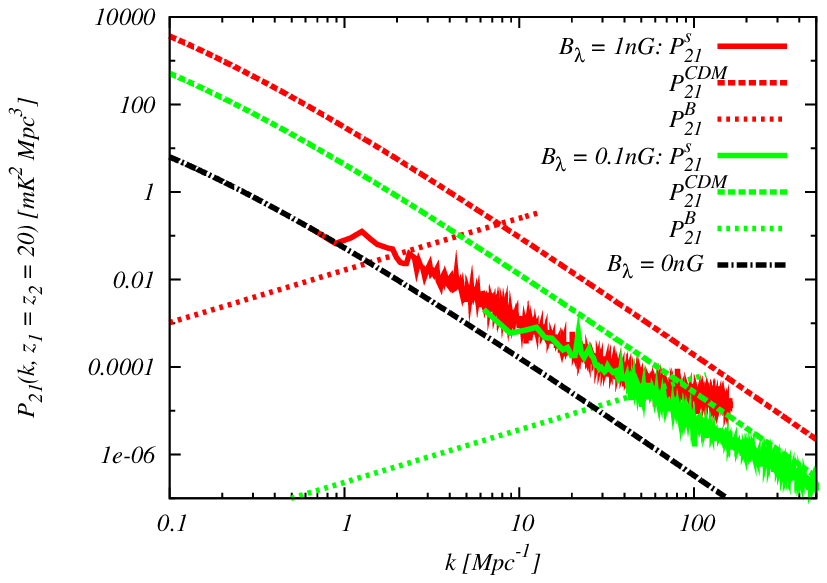}}
\end{minipage}
\begin{minipage}[m]{0.49\textwidth}
\rotatebox{0}{\includegraphics[width=1.0\textwidth]{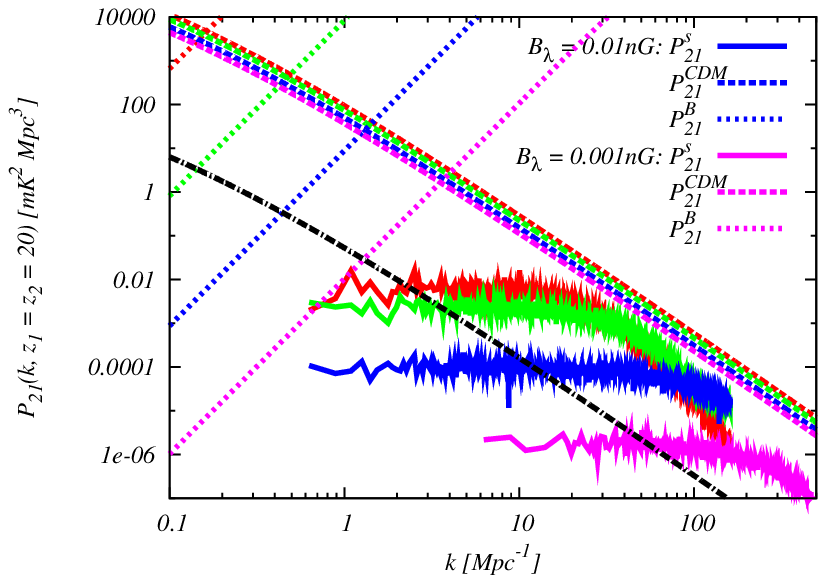}}
\end{minipage}
\caption{Power spectra of the 21cm fluctuations for $z_1 = z_2 =
 20$ when $n_B = -2.9$ (left) and $0$ (right). Here,
 $B_\lambda$ varies from 0 to 1 nG. For comparison, we separately plot
 the contributions from the temperature fluctuations ($P^s_{21}$: solid line), the standard matter fluctuations ($P^{\rm CDM}_{21}$: dashed line), and the magnetized matter fluctuations ($P^{B}_{21}$: dotted line). Note that we here do not plot the 0.01 and 0.001 nG cases when $n_B = -2.9$, and the lines of $P^s_{21}$ are broadened because it is one realization of the Monte Carlo simulations~(also in Fig.~\ref{fig:Cl21}).}
\label{fig:Pk21}
\end{figure}

Cross and auto power spectra of 21cm fluctuations are expressed as 
\begin{eqnarray}
\Braket{\prod_{i=1}^2 \delta T_{21}({\bf k_i}, z_i)} 
 \equiv (2\pi)^3 P_{21}(k_1, z_1, z_2)  \delta^{(3)}({\bf k_1} + {\bf k_2}) ~.\label{eq:Pk21}
\end{eqnarray}
The power spectra are directly estimated from the simulations.
In Fig.~{\ref{fig:Pk21}} we show the power spectra at $z_1 = z_2 =20$ for
magnetic field strengths ranging from $B_\lambda = 0$ nG to $1$ nG,
with the spectral index $n_B=-2.9$ and $0$.

The power spectrum of 21cm brightness temperature mainly has two
parts. The one is coming from the temperature fluctuation term
that is proportional to 
$\delta\left(x_{\rm HI}(T_s - T_\gamma)/ T_s\right ) \bar{n}_H$, and the other is the density fluctuation term proportional to
$\bar{x}_{\rm HI}(\bar{T}_s-\bar{T}_\gamma)/\bar{T}_s \delta n_H$, where $x_{\rm HI} = n_{\rm HI} / n_H = 1-x_i$ is the neutral fraction and the bar 
means the average value. The former contribution to the power spectrum is
estimated from the Monte Carlo simulations described earlier because of
the complicated nonlinearity and denoted by $P^{s}_{\rm 21}$, while
the latter can be estimated using a publicly available CMB code such as CAMB \cite{Lewis:2007kz} and
denoted by $P^{\rm mat}_{\rm 21}$. Note that the latter contribution can be further divided into two contributions under the presence of magnetic fields.
The Lorentz force acts on baryons and alters its density perturbations,
while dark matters are indirectly affected by magnetic fields via the
gravitational interaction with baryons. After cosmological
recombination, an evolution equation for total
matter fluctuations composed of baryon and dark matter fluctuations 
reads \cite{1996ApJ...468...28K,2005MNRAS.356..778S}
\begin{eqnarray}
\frac{d^2 \delta_m}{d t^2} = -2 H \frac{d \delta_m}{dt} + 4 \pi G \bar{\rho}_m
 \delta_m + \frac{\nabla \cdot [ (\nabla \times {\bf B}) \times {\bf B}]
 }{4 \pi \bar{\rho}_m a^2} ~,
 \label{eq:den-evo-w-mag}
\end{eqnarray}
where $\bar{\rho}_m$ is the background total matter energy density and $\delta _m $ is the
density contrast of them.
Solving Eq.~(\ref{eq:den-evo-w-mag})
with the assumption that there is no correlation between primordial
magnetic fields and primordial density fluctuations,
the density matter power spectrum can be divided into two parts as
\begin{equation}
P_m(k)  = P_m ^{\rm CDM} (k)+ P_m ^B(k), 
\end{equation}
where the first term $P_m ^{\rm CDM}(k)$ is originated from the primordial density
fluctuations which are exactly the same as those in the standard $\Lambda$CDM model.
The second term $P_m^B$ represents the power spectrum of the density fluctuations produced by
primordial magnetic fields which depends on the power spectrum of the magnetic fields, as shown in
Ref.~\cite{2012PhRvD..86d3510S}.
According to Eqs.~(\ref{eq:diff_Tb}) and (\ref{eq:opt}),
the power spectrum of 21cm
fluctuations due to the density fluctuations can be given by
\begin{equation}
 P^{\rm mat}_{21}(k) =  \delta \overline{T}_{21} ^2 
\left[ P_m ^{\rm CDM} (k)+ P_m ^B(k) \right] \equiv P^{\rm CDM}_{21}(k) + P^B_{21}(k) ~,
\end{equation}
where $\delta \overline{T}_{21} $ is the mean differential brightness
temperature obtained from Eqs.~(\ref{eq:diff_Tb}) and (\ref{eq:opt}) with
the background density $\bar{n}_{\rm HI}$ and the mean spin temperature
$\bar{T}_s$.\footnote{Although we ignore all cross-correlation terms, the cross correlation between the temperature
and density fluctuations induced by magnetic fields can
also contribute to the power spectrum $P_{21}$.}

In Fig.~{\ref{fig:Pk21}}, we separately plot these three contributions. We
find that magnetic fields with nanogauss levels significantly enhance the power
over the wide range of scales through the density fluctuation term $P^{\rm mat}_{\rm 21}$,
because they realize that 
${(\bar{T}_s-\bar{T}_\gamma)/\bar{T}_s}_{\rm |PMF} \sim 1 \gg
{(\bar{T}_s-\bar{T}_\gamma)/\bar{T}_s}_{\rm |no~PMF}$, where the
subscripts PMF and no PMF, respectively, represent the values with and without primordial magnetic
fields, and give larger
density fluctuations especially on small scales. 
If we consider the case with $n_B = 0$, even weaker magnetic fields
with strength as small as $B_\lambda = 10^{-3}$ nG can amplify the
standard signal with no primordial magnetic fields (black dot-dashed line) by 3 orders of magnitude (see the right panel in the figure). 

The contributions from the temperature fluctuations, {$P^{s}_{21}$}, are always subdominant for magnetic fields with
a bluer spectrum, as shown in the right panel in
Fig~{\ref{fig:Pk21}}. However, the density fluctuations due to magnetic
fields are suppressed below the magnetic Jeans scale. In the figure, to
take into account this suppression, we introduce cutoffs to 
 the contributions from $P_m^B$, namely, $P_{21}^B$, by hand at the magnetic Jeans scales.
We find that $P_{21}^s$ can give a comparable
contribution with the primordial density fluctuation term, $P_{21}^{\rm
CDM}$, only on scales smaller than the cutoff scales for nearly scale-invariant
magnetic fields ($n_B=-2.9$; the left panel in the figure).

According to Eq.~(\ref{eq:gamma}), as the magnetic field amplitude $B_\lambda$ increases,
the heating rate due to the ambipolar diffusion becomes large. However,
when the gas temperature reaches $T_g \sim 3000~$K, the
temperature no longer rises and, instead, the ionization fraction grows.
This ionization fraction growth suppresses the dissipation rate due to
ambipolar diffusion as shown Eq.~(\ref{eq:gamma}). As a result, the gas
temperature remains $T_g \sim 3000~$K and the ionization fraction gradually
grows until the cosmological expansion term dominates in Eq.~(\ref{eq:Te}).
During this regime, the fluctuations of the gas temperatures start
to saturate, because the local gas temperature reaches $3000~$K at many
different places.
Therefore, when $B_\lambda$ is large~($B_\lambda > 0.1~$nG), the dependence of the $P^s_{21}$ amplitude on $B_\lambda$ becomes milder 
than in the cases with small $B_\lambda$
as shown in Fig.~\ref{fig:Pk21}.

%%%%%%%%%%%%%%%%%%%%%%%%%%%%%%%%%%%%%%%%%%%%%%%%%%%%%%%%%%
\section{Angular power spectrum of brightness temperature fluctuations} \label{sec:cl}

One of the main aims for 21cm observations is to measure the angular
power spectrum (or three-dimensional power spectrum) of the 21cm fluctuations at each
redshift. In this section, we evaluate the angular power spectra from
the 21cm maps obtained in the previous section.  

The 21cm fluctuations projected on a spherical shell at $z = z_*$ with its width $\Delta z_*$ are expressed as
\begin{eqnarray}
\delta T_{21}(\hat{\bf n}, z_*, \Delta z_*) = \int_0^{\chi_\infty} d\chi W(\chi, \chi_*, \Delta \chi_*) \delta T_{21}({\bf x}),
\end{eqnarray}
where $\chi(z)$ denotes the conformal distance, $\chi_* \equiv
\chi(z_*)$, $\chi_\infty \equiv \chi(\infty)$, $\Delta \chi_* \equiv
\chi(z_*+\frac{\Delta z_*}{2}) - \chi(z_*-\frac{\Delta z_*}{2})$, and
$\hat{\bf n} \equiv {\bf x} / \chi_* $. A normalized window function
$W(\chi, \chi_*, \Delta \chi_*)$ is
associated with the bandwidth of an observation.
Generally, the window function is a function of the frequency
centered at the observed frequency. However, because there is one to one correspondence between the frequency and the
conformal distance, 
we here adopt the following Gaussian function for simplicity, 
\begin{eqnarray}
W(\chi, \chi_*, \Delta \chi_*) = \frac{1}{\sqrt{2 \pi ({\Delta \chi_*}/{2})^2}} 
\exp\left[- \frac{(\chi - \chi_*)^2}{2 ({\Delta \chi_*}/{2})^2}\right] ~. \label{eq:window}
\end{eqnarray}
In this section, we shall analyze the angular power spectrum of $\delta T_{21}(\hat{\bf n}, z_*, \Delta z_*) $ originating from primordial magnetic fields. 

An expression in multipole space reads 
\begin{eqnarray}
\delta T_{21, \ell m}(z_*, \Delta z_*) = \int d^2 \hat{\bf n}~ Y_{\ell m}^* (\hat{\bf n}) \delta T_{21}(\hat{\bf n}, z_*,\Delta z_* )~.
\end{eqnarray}
Under the assumption that 21cm signals are statistically isotropic,
their angular power spectrum does not depend on $m$, 
\begin{equation}
\Braket{\prod_{i=1}^2 \delta T_{21, \ell_i m_i}(z_*, \Delta z_*)} 
= (-1)^{m_1} \delta_{\ell_1 \ell_2} \delta_{m_1, -m_2} C_{\ell_1}(z_*, \Delta z_*) ~, \\ 
\end{equation}
and $C_\ell$ is obtained by
\begin{equation}
C_{\ell}(z_*, \Delta z_*) = \frac{2}{\pi} \int_0^\infty k^2 dk 
\left[ \prod_{i=1}^2 \int_0^{\chi_\infty} d\chi_i W(\chi_i, \chi_*, \Delta \chi_*) 
j_{\ell}(k \chi_i) \right]
P_{21}(k, z(\chi_1), z(\chi_2)) 
  ~. \label{eq:all_cl}
\end{equation}
For computations on small scales, a reduced formula under the flat-sky
coordinate, namely, $\hat{\bf n} \to ({\boldsymbol \theta}, 1)$, is
useful. The representations of $\delta T_{21}$ and the angular power
spectrum in ${\boldsymbol \ell}$ space are given by 
\begin{eqnarray}
\delta T_{21}({\boldsymbol \ell}, z_*, \Delta z_*) &=& \int d^2
 {\boldsymbol \theta} e^{- i {\boldsymbol \ell} \cdot {\boldsymbol
 \theta}} \delta T_{21} (\hat{\bf n}, z_*, \Delta z_*) ,
 \nonumber \\
\Braket{\prod_{i=1}^2 \delta T_{21}({\boldsymbol \ell_i}, z_*, \Delta z_*)} 
&=& (2\pi)^2 \delta^{(2)} ( {\boldsymbol \ell_1} + {\boldsymbol \ell_2} ) C(\ell_1, z_*, \Delta z_*) ~. 
\end{eqnarray}

Applying the so-called Limber approximation that Fourier waves along the $z$~axis cancel each other out for $\lambda \sim 1/k_z \ll \chi / \ell$
yields an expression for the angular power spectrum~\cite{2003moco.book.....D}:
\begin{eqnarray}
C (\ell, z_*, \Delta z_*) &=& \int_0^{\chi_\infty} d\chi \frac{W^2 (\chi, \chi_*, \Delta \chi_*)}{\chi^2} P_{21} \left(\frac{\ell}{\chi}, z(\chi), z(\chi)\right) ~.
\label{eq:flat_limb_cl}
\end{eqnarray}
This is in good agreement with the exact formula (\ref{eq:all_cl}) when
the cancellations of Fourier waves happen frequently within the width of
the window function, namely, $\ell \gg {\chi_*}/{\Delta \chi_*}$. The
following numerical results focusing on such small scales are estimated
by Eq.~(\ref{eq:flat_limb_cl}) because this enforces many fewer numerical
operations than Eq.~(\ref{eq:all_cl}).

\begin{figure}
 \begin{minipage}[m]{0.49\textwidth}
\rotatebox{0}{\includegraphics[width=1.0\textwidth]{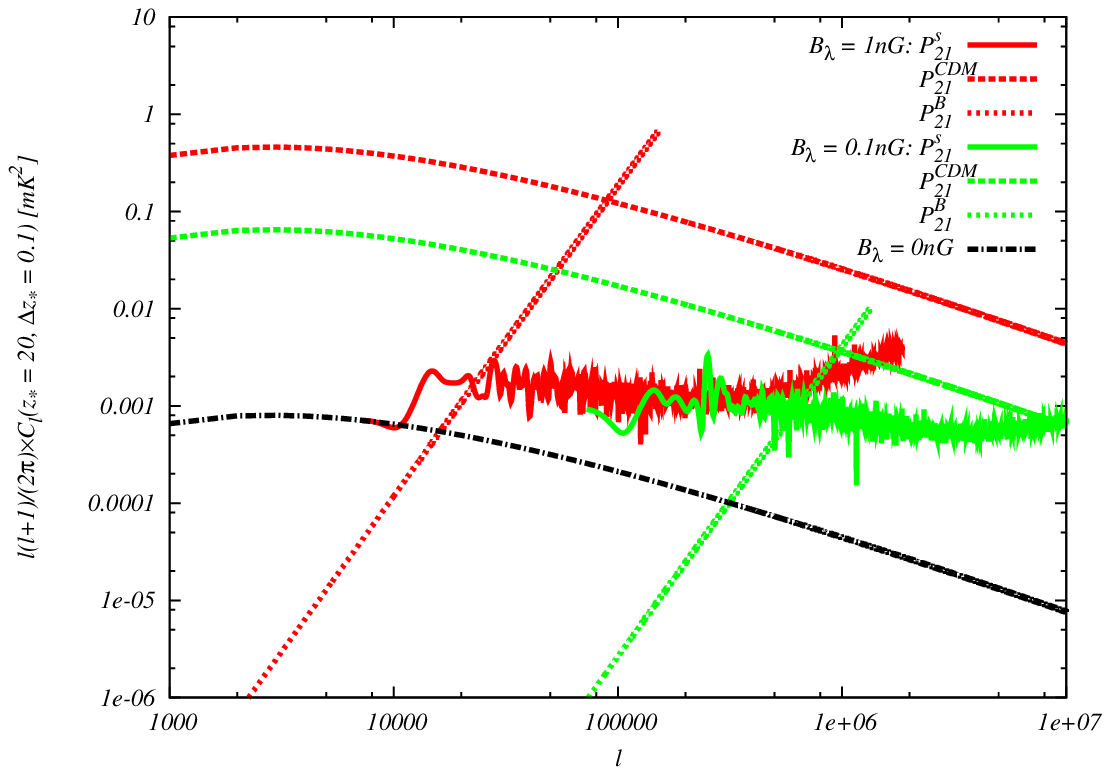}}
\end{minipage}
\begin{minipage}[m]{0.49\textwidth}
\rotatebox{0}{\includegraphics[width=1.0\textwidth]{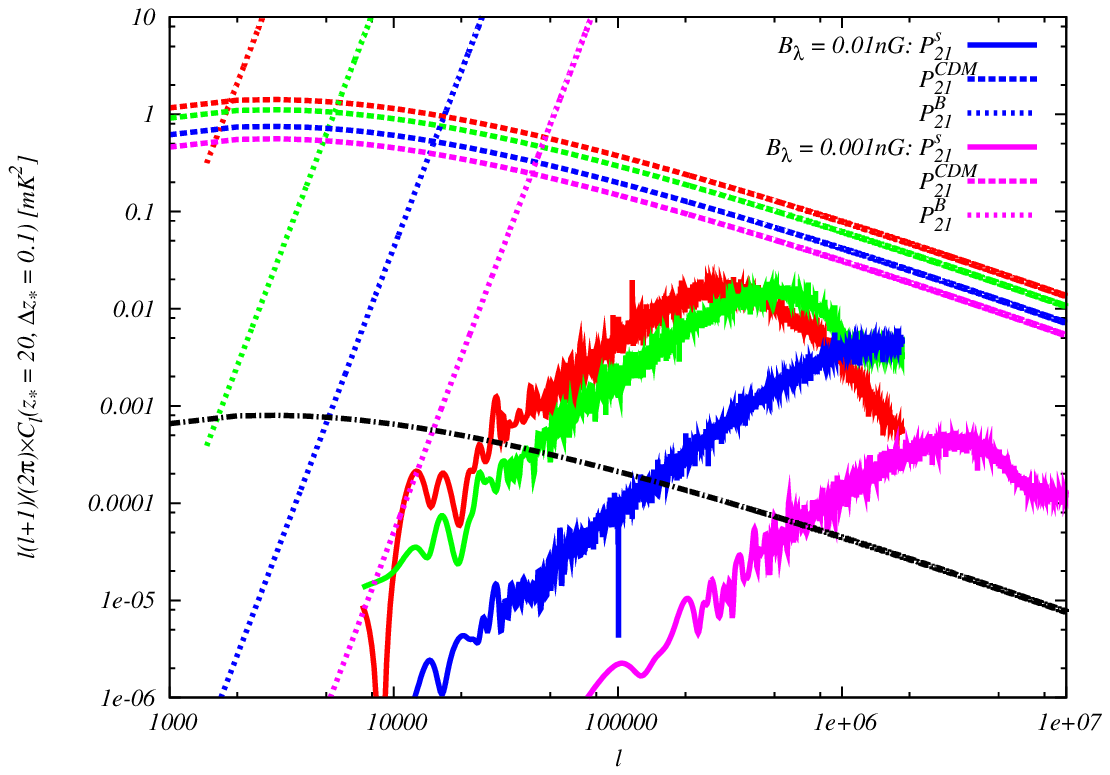}}
\end{minipage}
\caption{Angular power spectra of the 21cm fluctuations for $z_* = 20$ and $\Delta z_* = 0.1$ when $n_B = -2.9$ (left) and $0$ (right). The settings are identical to Fig.~\ref{fig:Pk21}.}
\label{fig:Cl21}
\end{figure}

We plot the angular power spectra in Fig.~\ref{fig:Cl21}.
Similar to Fig.~\ref{fig:Pk21}, we separately show the three
contributions, the temperature fluctuations due to the ambipolar
diffusion $P^s_{21}$, the primordial density fluctuations $P^{\rm CDM}_{21} = \delta \overline{T}_{21} ^2 P_m^{\rm CDM}$, and the density fluctuations induced by magnetic fields $P^B_{21} = \delta \overline{T}_{21} ^2 P_m^{B}$. Although $P_m^{\rm CDM}$ is independent from 
magnetic fields, the amplitude of $P^{\rm CDM}_{21}$ is sensitive to
$B_\lambda$ because magnetic fields increase the spin temperature
through heating the background gas temperature due to the ambipolar diffusion. 

For the power-law spectrum $P_{21} \propto k^n$,
Eq.~(\ref{eq:flat_limb_cl}) tells us that the angular power spectrum
$\ell^2 C_\ell$ is proportional to $\ell^{n+2}$.
Therefore, Fig.~\ref{fig:Cl21} shows that $\ell^2 C_\ell$ due to $P^s_{21}$
is proportional to $\ell^{-0.1}$ and $\ell^2$ for $n_B=-2.9$ and $0$,
respectively, because the spectral index $n$ of $P^s_{21}$ is roughly $n = -2.1$ for $n_B=-2.9$ and $n = 0$ for $n_B = 0$, as shown in Fig~\ref{fig:Pk21}. We can also see that $k$ space signatures are reflected in $\ell$ space by following $\ell \sim k \chi_*$ with $\chi_* \sim 12$ Gpc. 

As seen in Fig.~\ref{fig:Pk21}, $P^s_{21}$ is dominated by $P^B_{21}$. However,  
on smaller scales than the magnetic Jeans scale, 
$P^B_{21}$ is expected to be strongly suppressed and the temperature fluctuations due to the ambipolar diffusion can significantly contribute
to the 21cm signals with the primordial density fluctuation
contributions, especially in the case with nearly scale-invariant
magnetic fields ($n_B=-2.9$; the left panel in the figure).

%%%%%%%%%%%%%%%%%%%%%%%%%%%%%%%%%%%%%%%%%%%%%%%%%%%%%%%%%%
\section{Conclusion}

In this paper, we have studied cosmological 21cm signals induced by
primordial magnetic fields, focusing on the ambipolar diffusion
of magnetic fields.
The ambipolar diffusion heats the gas temperature and the heating
rate depends on the magnetic field strength. Therefore, when primordial
magnetic fields are tangled, the ambipolar diffusion not only increases
the background gas temperature, but also generates the fluctuations of the
gas temperature. These fluctuations alter the fluctuations of
cosmological 21cm lines. 
We have evaluated these fluctuations due to the ambipolar diffusion,
calculating the thermal evolution of the hydrogen gas with Monte Carlo
simulations.
We have shown that the 21cm fluctuations due to the ambipolar diffusion
depend on the magnetic field properties such as the strength and the spectral
index of magnetic fields. We have also found that the fluctuations start
to saturate for the magnetic field strength $B_\lambda > 0.1~$nG.
This is because the gas temperature cannot increase beyond $\sim 3000$~K due to the balance
between the ionization fraction and the ambipolar diffusion rate.
The gas temperatures in most regions reach this critical temperature for
large $B_\lambda$ and, as a result, the fluctuations cannot be amplified. 

Primordial magnetic fields can give the other two effects on the 21cm
fluctuations, as discussed in Refs.~\cite{2006MNRAS.372.1060T,2009ApJ...692..236S}.
One is the amplification of the 21cm fluctuations originating from the
primordial density fluctuations because the dissipated magnetic field
energy increases the background gas temperature.
The other is the additional density fluctuations which are generated by
primordial magnetic fields after the epoch of recombination.
We have compared these two contributions with
the contribution from the temperature fluctuations obtained by our simulations.
Our result has shown that 
the contributions of the temperature fluctuations are subdominant on
observation scales of future observations such as SKA.
On these  scales, the most important effect of primordial magnetic fields is the amplification due to the heating of the background gas temperature.
This result is consistent with Refs.~\cite{2006MNRAS.372.1060T,2009ApJ...692..236S}. The temperature fluctuation contribution can give the non-negligible contribution only on small scales~($\ell >10^5$) for nearly scale-invariant magnetic fields.

In this paper, we have focused on the effect of primordial magnetic fields on 21cm
signals before the epoch of reionization.
However, near future observations such as SKA are planned to observe the 21cm radiation during the epoch of reionization.
In this paper, although we have taken into account the thermal ionization,
we have not considered the photoionization by first stars and
galaxies. The gas temperature heated by the dissipation of magnetic
field energy modifies the Jeans mass~\cite{2005MNRAS.356..778S,2008PhRvD..78h3005S}, and 
the additional density fluctuations due to primordial magnetic fields
enhance the abundance of ionization photon
sources~\cite{2006MNRAS.368..965T,2008PhRvD..78h3005S}.
These effects are expected to modify the 21cm fluctuations from those
in the standard $\Lambda$CDM model, and they can give a significant contribution to observable 21cm fluctuations. Therefore, to study the feasibility of the constraint on primordial magnetic fields by future observations, a detailed evaluation of these effects during the epoch of reionization is important.
We leave this issue for a future work.

%%%%%%%%%%%%%%%%%%%%%%%%%%%%%%%%%%%%%%%%%%%%%%%%%%%%%%%%%
\acknowledgments 
M. S. is supported in part by a Grant-in-Aid for JSPS Research under Grant No.~25-573 and the ASI/INAF Agreement I/072/09/0 for the Planck LFI Activity of Phase E2. H. T. is supported by the DOE at Arizona State University. K. I. is also supported by Grant-in-Aid No. 24340048 from the Ministry of Education, Sports, Science and Technology  of Japan
%\appendix*
%\section{appendix}
%%%%%%%%%%%%%%%%%%%%%%%%%%%%%%%%%%%%%%%%%%%%%%%%%%%%%%%%%%%%%%%%%

\bibliography{paper}

\end{document}